\documentclass[aps,prd,showpacs,preprintnumbers]{revtex4}
\usepackage{graphicx}
\usepackage{epsfig}
\usepackage{amsmath}
\usepackage{subfigure}

\def \beq{\begin{equation}}
\def \eeq{\end{equation}}
\def \beqa{\begin{eqnarray}}
\def \eeqa{\end{eqnarray}}
\def \beal{\begin{align}}
\def \enal{\end{align}}
\begin{document}

\title{Chiral phase transition with mixing between scalar quarkonium and tetraquark}

\bigskip
\bigskip
\author{Tamal K. Mukherjee$^{1,2}$}
\email{mukherjee@ihep.ac.cn}
\author{Mei Huang$^{1,2}$}
\email{huangm@ihep.ac.cn}
\affiliation{$^1$  {Institute of High Energy Physics, Chinese Academy of
Sciences, Beijing, China} \\
$^2$ {Theoretical Physics Center for Science Facilities, Chinese Academy of
Sciences, Beijing, China} }
\date{\today }
\bigskip

\begin{abstract}
In the framework of two-flavor extended linear sigma model with mixing
between scalar quarkonium and tetraquark, we investigate the role of
the tetraquark in the chiral phase transition. We explore various scenarios 
depending on the value of various parameters in our model. The physical mass spectrum of mesons put a tight constraint on the parameter set of our model. We find a sufficiently strong cubic self interaction of the tetraquark field can drive the chiral phase transition to first order even at zero quark chemical potential. Weak or absence of the cubic self interaction term of the tetraquark field make the chiral phase transition crossover at vanishing density.
\end{abstract}

\maketitle

\section{Introduction}
The role of chiral condensate is well known and well studied in
the context of chiral phase transition. Recently, the possible role of
tetraquark condensate in connection to the chiral phase transition
is also being considered~\cite{heinz, harada}. The reason behind such
consideration stems from the unsettled nature of the lightest scalar
$f_0(600)$ ($f_0(500)$ in \cite{Pelaez:2013lya}) or $\sigma$ meson.
This issue is part of the
unresolved nature of the scalar mesons below 2 ${\rm GeV}$. There are about
19 scalar resonances found below 2 ${\rm GeV}$ which cannot be explained by
the naive quark model. Their mass spectrum and decay patterns are also
quite contrary to what is expected from the quark model. An intense effort
is going on to understand the nature and properties of these mesons
(see refs.~\cite{pdg2010,Scalar, Scalar-lightGB,Glueball-Review} and
references therein).

Theoretical understanding of the lightest scalar $\sigma/f_0(600)$ is
important as it is believed to be the Higgs Bosons of QCD and plays an
important role in chiral symmetry breaking. Though its existence has been
confirmed from the $\pi \pi$ scattering process \cite{Caprini:2005zr,Scalar},
the consensus on its nature is still elusive. Conventionally $f_0(600)$ is
regarded as composed of quark-antiquark. But in order to solve the mass
hierarchy problem for scalar mesons below 1 GeV, Jaffe~\cite{jaffe} in 1977
proposed to consider the scalar meson below 1 ${\rm GeV}$ as tetraquark states
and those above 1 ${\rm GeV}$ to be quarkonium states. Thus, in this
picture $f_0(600)$ is predominantly a tetraquark states whereas $f_0(1370)$ is
the lightest quarkonium state made up of quark-antiquark. The sizable tetraquark
component has also been demonstrated in a recent Lattice simulation study~\cite{Prelovsek:2010kg}. However, there are other suggestions as well,
for example, recent data from $\pi \pi$ and $\gamma \gamma$ scattering \cite{Minkowski:1998mf,arXiv:0804.4452}, the
K-matrix analysis \cite{Mennessier:2010xg} suggests that it has sizable
glueball content.

The role of chiral condensate as an order parameter for chiral phase transition
is well established. But, the role of tetraquark condensate is not understood
and work in this direction has recently been started~\cite{heinz, harada}.

In~\cite{heinz} the implications of mixing between tetraquark and quarkonium
fields on chiral phase transition is studied for zero baryon chemical
potential. They favor the scenario where, $f_0(600)$ is tetraquark dominated
and the heavy $f_0(1370)$ is quarkonium dominated. According to their study,
the order of the phase transition is strongly correlated with the extent of
mixing between the two fields. For a weak coupling constant for the mixing
term, a soft first order phase transition is obtained. On the other hand for
a strong coupling constant for the mixing term give rise to a crossover
transition. Moreover, the most important and interesting result coming
out from their study is that beyond a certain maximum temperature the
nature of the heavy and lighter mesons is exchanged. The heavy $f_0(1370)$
becomes tetraquark dominated and the lighter $f_0(600)$ turns into quarkonium
dominated and becomes degenerate with the pion after the chiral symmetry
restoring phase transition.

Whereas in~\cite{harada}, an alternate breaking of chiral symmetry in dense matter
was proposed. Using Ginzburg-Landau effective potential consisting of two and
four quark states they show in dense matter a possible phase may arise
where chiral symmetry is spontaneously broken but its center symmetry
remains unbroken. In this phase conventional chiral condensate vanishes and
the chiral symmetry breaking is due to the presence of quartic condensate.
Finally chiral symmetry is restored as quartic condensate also vanishes.
Existence of a tricritical point is also predicted between broken and
unbroken center symmetric phase. Thus in this scenario, restoration
of chiral symmetry occurs in two steps.

These studies warrant us to study the effect of mixing between the
quarkonium and tetraquark condensates on the chiral phase transition
in detail. Here in this work, we study two flavor chiral phase
transition within the framework of extended linear sigma model taking into
account both quarkonium and tetraquark effective fields. We fix
parameters from the physical meson masses. Depending on the possible
values of the various parameters, the resulting phase diagram is
discussed.

The paper is organized as follows: in the next section we discuss about 
the model we are going to consider and how the various parameters
in the model is fixed. In section III we present our result and finally
we summarize and conclude in the last section.

\section{The model}

We are going to investigate the effect of quarkonium and tetraquark
mixing on chiral phase transition in the framework of quark-meson
model. In this model, quarks
propagate in the background potential of mesonic fields and interact
with the vacuum expectation values of the scalar mean fields via Yukawa
coupling. The generic form of the Lagrangian consist of a fermionic part
($\cal{L}_{\text{q}}$) and a mesonic field part ($\cal{L}_{\text{m}}$)
and can be written as:
\begin{eqnarray}
\cal{L} &=& \cal{L}_{\text{q}} + \cal{L}_{\text{m}} \\ \nonumber
&= & {\bar q} (i\gamma^{\mu}\partial_{\mu} - g_3 \Phi -g_4 \Phi^\prime ) q
+ \cal{L}_{\text{m}},
\label{lagB}
\end{eqnarray}
with the mesonic part of the Lagrangian: 
\begin{eqnarray}
\cal{L}_{\text{m}} &= & {\rm Tr}(\partial_\mu \Phi \partial^{\mu}
\Phi^{\dagger})+ {\rm Tr}(\partial_\mu \Phi^\prime \partial^{\mu}
\Phi^{\dagger \prime})
 -{m_\Phi}^2 {\rm Tr} (\Phi^\dagger \Phi)
-{m_{\Phi^\prime}}^2 {\rm Tr} (\Phi^{\dagger \prime} \Phi^\prime) \nonumber \\
  & & + \frac{\lambda_1}{2} {\rm Tr} (\Phi^\dagger \Phi \Phi^\dagger \Phi) \notag
  + \frac{\lambda_2}{2}
 {\rm Tr} (\Phi^{\dagger \prime} \Phi^\prime \Phi^{\dagger \prime} \Phi^\prime) \nonumber \\
 & & + g_2 {\rm Tr} ( \Phi^\prime \Phi^\prime \Phi^\prime)
 - g_1 {\rm Tr} (\Phi^\prime) {\rm Tr} (\Phi) {\rm Tr} (\Phi) \nonumber \\
& & +k [{\rm Det}(\Phi) + h.c.] -h [{\rm Tr} (\Phi) + h.c.].
\label{lagM}
\end{eqnarray}
Where, for two light flavors the quark field "q" can be represented as
$q = (u, d)$ and $g_3$, $g_4$ are the Yuakwa coupling constants for the
quarkonium and tetraquark fields respectively. The mesonic Lagrangian
part has two effective fields: a $2\times2$ matrix field $\Phi$ which denotes
the bare quarkonium field and a $2\times2$ matrix field $\Phi^\prime$ which
denotes the bare tetraquark field. Following the convention of the linear sigma
model, we express the quarkonium and the tetraquark fields as:
\begin{eqnarray}
\Phi & = & \frac{1}{2} (\sigma_b + \eta_b) + \frac{1}{2} (\vec{\alpha_b}
 +i \vec{\pi_b}).\vec{\tau},  \\
\Phi^\prime &= & \frac{1}{2} (\sigma_b^\prime + \eta_b^\prime) + \frac{1}{2}
(\vec{\alpha_b^\prime}  +i \vec{\pi_b^\prime}).\vec{\tau},
\end{eqnarray}
with $\tau_i$ (i = 1, 2, 3) represents $2 \times 2$ Pauli matrix.
The transformation properties of these
fields under $U(2)_L \times U(2)_R$ symmetry are defined as follows:
\begin{eqnarray}
\Phi & & \rightarrow U_L \Phi {U^\dagger}_R,  \\
\Phi^\prime & & \rightarrow  U_L \Phi^\prime {U^\dagger}_R\,\,,
\end{eqnarray}
where $U_{L,R}$ are group elements of the $U(2)_L \times U(2)_R$
symmetry.

The mesonic spectra consist of sixteen physical mesons: pair of scalar isoscalar
$\{ f_0(600), f_0(1370) \}$, pair of pseudoscalar isoscalar $\{ \eta_p,
\eta_p^\prime \}$, pair of scalar isovector $\{ \vec{\alpha_p},
\vec{\alpha_p}^\prime \}$
and pair of pseudoscalar isovector $\{ \vec{\pi_p}, \vec{\pi_p}^\prime \}$.
Here, pseudoscalar isoscalar $\eta_p$ and $\eta_p^\prime$ mesons are composed
of $u$ and $d$ quarks only. The bare quarkonium and tetraquark fields mixed with
each other to give rise to physical mesonic fields: one of them being
quarkonium dominated and the other tetraquark dominated mesons.

In the mesonic part Lagrangian Eq.(\ref{lagM}), the cubic term for the tetraquark
meson with the coupling constant $g_2$, the mixing term
between quarkonium and tetraquark with the coupling constant $g_1$ and the last term
mimicking the finite quark mass for the quarkonium with coupling constant $h$ explicitly
breaks the $U(2)_L \times U(2)_R$ symmetry. Whereas, the instanton determinant term
explicitly break the axial $U(1)_A$ symmetry. The other terms in the potential part
of the Lagrangian are invariant under $U(2)_L \times U(2)_R$ symmetry. However, we
spontaneously break the $SU(2)_A$ part of the symmetry of these terms as well by assuming
vacuum expectation values for the $\sigma_b$ and $\sigma_b^\prime$ fields.
The mass and the quartic interaction terms are the standard terms used
in the linear $\sigma$ model. An explicit symmetry breaking term to
account for the finite quark mass and an instanton determinant term for
the field $\Phi$ are also used. The explicit chiral symmetry breaking terms
for the field $\Phi$ renders $\pi_b$ and $\eta_b$ massive. The instanton
determinant term is responsible for the splitting of masses between
$\pi_b$ and $\eta_b$. The choice of the cubic term
is motivated from the study in ref.~\cite{harada}.

We will investigate the chiral phase transition at the mean field level.
Following the standard procedure, we expand the fields around the vacuum
expectation values:
$\sigma_b = \sigma + \sigma_f$ and $\sigma_b^\prime = \chi + \sigma_f^\prime$.
Where, $\sigma$ and $\chi$ are the vacuum expectation values of the corresponding
fields. Keeping only the mean fields and integrating out the fermionic fields we
obtain (neglecting ultraviolet divergent vacuum energy term \cite{scavenius,schaefer}) 
the expression for the thermodynamical potential at temperature $T$ and chemical potential 
$\mu$ as:
\begin{eqnarray}
  \Omega & =  &  U (\sigma, \chi)  - 2 T  N_c N_f\int \frac{d^3 q}{(2 \pi)^3} \nonumber \\
& & [{\rm ln}(1 + e^{-(E_q - \mu)/T}) + {\rm ln}(1 + e^{-(E_q + \mu)/T})],
\end{eqnarray}
where
\begin{eqnarray}
  U (\sigma, \chi)&  = & - \frac{1}{2} {m_\Phi}^2 \sigma^2
  - \frac{1}{2} {m_{\Phi^\prime}}^2
\chi^2 +\frac{1}{16} \lambda_1 \sigma^4
+\frac{1}{16} \lambda_2 \chi^4 \nonumber \\
& & +\frac{1}{4} g_2 \chi^3 - g_1 \sigma^2 \chi
+\frac{1}{2} k \sigma^2 - 2 h \sigma.
\end{eqnarray}
The single particle energy is given by $E_q = \sqrt{p^2 +{m_q}^2}$ and
the constituent quark mass ($m_q$) is given by $m_q = g_3 \sigma + g_4 \chi$.
The number of colours $N_c$ and flavors $N_f$ of quark used in this paper
are 3 and 2 respectively.

From the extremum condition of the thermodynamic potential we obtain equation of
motions for $\sigma$ and $\chi$:
\begin{equation}
\frac{\partial \Omega}{\partial \sigma} = 0, \hskip 0.1in
\frac{\partial \Omega}{\partial \chi} = 0.
\label{min}
\end{equation}

We will solve this set of coupled equation of motions Eq.(\ref{min})
self consistently at each values of
temperature $T$ and chemical potential $\mu$ to determine the behavior of $\sigma$
and $\chi$ as a function of temperature and chemical potential, and analyze the effect
of quarkonium-tetraquark mixing on the chiral phase transition.

\section{Parameter Fixing in the vacuum}

There are total 12 parameters in our model: ${m_\Phi}^2$,
$m_{\Phi^\prime}^2$, $\lambda_1$, $\lambda_2$, $g_1$,
$g_2$, $k$, $h$, $g_3$, $g_4$ and zero temperature values
of $\sigma$, $\chi$. Out of
these 12 parameters, the coupling constants $g_3$,
$g_4$ are from the fermionic part ($\cal{L}_{\text{q}}$) of the
Lagrangian and the other 10 are from the mesonic part
($\cal{L}_{\text{m}}$).
The values of the 10 parameters in the mesonic part of the
Lagrangian are determined from the physical meson masses,
the pion decay constant ($f_\pi = 92.4$ MeV) and
two extremum conditions for the mesonic potential:
\begin{eqnarray}
\frac{\partial U (\sigma, \chi)}{\partial \sigma} = 0,
\hskip 0.1in
\frac{\partial U (\sigma, \chi)}{\partial \chi} = 0.
\label{minU}
\end{eqnarray}

Values of the parameters so fixed are kept constant
for the whole range of temperature and chemical
potential. The physical meson masses are so chosen
that for each kind of meson, one of the mass is below
$1 {\rm GeV}$ and the other is above $1 {\rm GeV}$. One
of them is likely choice for the quarkonium dominated
meson and the other is tetraquark dominated as found by other studies~\cite{fariborz,tamal}.

The physical meson masses are obtained by diagonalizing
the bare meson mass matrices. The expression for those
bare matrices as a function of quarkonium, tetraquark
fields and the coupling constants, are noted below:

For the sigma mesons we have:
\begin{align}
\left(  M_{f_0}^2  \right) =
 \left[
  \begin {array}{cc}
\frac{1}{2} \, \lambda_1 \, \sigma^2 \, + 2 \, \frac{h}{\sigma}
&
-2 \, g_1 \, \sigma
\\
-2 \, g_1 \, \sigma
&
\frac{1}{2} \, \lambda_2 \, \chi^2 \, + \frac{3}{4} \, g_2 \, \chi
\, + g_1 \, \frac{\sigma^2}{\chi}
\end{array}
 \right].
\end{align}

For pions we have:
\begin{align}
\left(  M_\pi^2  \right) =
 \left[
  \begin {array}{cc}
2 \, g_1 \, \chi + 2 \, \frac{h}{\sigma}
&
0
\\
0
&
g_1 \, \frac{\sigma^2}{\chi} \, - \frac{9}{4} \, g_2 \, \chi
\end{array}
 \right].
\end{align}

For eta we have,
\begin{align}
\left(  M_\eta^2  \right) =
 \left[
  \begin {array}{cc}
4 \, g_1 \, \chi - 2 \, k \, + 2 \, \frac{h}{\sigma}
&
2 \, g_1 \, \sigma
\\
2 \, g_1 \, \sigma
&
- \frac{9}{4} \, g_2 \, \chi \, + g_1 \, \frac{\sigma^2}{\chi}
\end{array}
 \right].
\end{align}

Lastly bare mass matrix for the $\alpha_p$ meson reads,
\begin{align}
\left(  M_\alpha^2  \right) =
 \left[
  \begin {array}{cc}
\frac{1}{2} \, \lambda_1 \, \sigma^2 \, + 2 \, \frac{h}{\sigma} \,
+ 2 \, g_1 \, \chi \, - 2 \, k
&
0
\\
0
&
\frac{1}{2} \, \lambda_2 \, \chi^2 \, + \frac{3}{4} \, g_2 \, \chi \,
+ g_1 \, \frac{\sigma^2}{\chi}
\end{array}
 \right].
\end{align}

From the mass matrices, we find that there is no mixing for the pion and
$\alpha$ mesons. We choose the
lightest pion as a quarkonium meson and the heavier counterpart as
tetraquark meson in its quark content. This is in agreement
with our current understanding. Since there is no mixing for pion, we
define the zero temperature value of $\sigma$ equal to the pion decay
constant ($\sigma = f_\pi$).
From the expression of pion mass we see the symmetry breaking terms
contribute towards its mass and the absence of those terms in our
Lagrangian would make pion massless. The same statement is also holds for the eta
mesons although in this case there is mixing between quarkonium and
tetraquark fields. In absence of mixing, $\left(  M_\eta^2  \right)_{11}$
would represent the physical eta meson mass and comparing it with
conventional pion mass $\left(  M_\pi^2  \right)_{11}$ we find the difference
between their masses are coming from the instanton term, which is in line
with our expectation.

In the following, we discuss three sets of parameters, which will be
used for our analysis of chiral phase transition in
Sec.\ref{sec:phasetransition}.

\subsection{Case I: $\lambda_2$, $g_2$, $k, h = 0$}
Here, in the simplest version of the model we want to explore the scenarios 
where the lowest scalar is either quarkonium dominated or tetraquark dominated 
meson. We find, within the limit of physical meson masses (including the experimental 
uncertainty of $m_{\pi^\prime}$: 1.2-1.4 GeV, $m_{f_0(600)}$: 0.4-1.2 GeV and 
$m_{f_0(1370)}$: 1.2-1.5 GeV), the scenario where lightest scalar isoscalar is 
tetraquark dominated meson cannot be realized. For this, in this 
case, we take the value of the physical meson masses slightly different from their 
real world values. To make comparison between the two scenarios meaningful, we keep 
the physical meson masses for both the scenarios as close as possible (see table~\ref{case1in}). 
In this case, the value of the parameters $\lambda_1$, $g_1$ and the
zero temperature value of $\chi$ are calculated using the physical
masses of $m_{f_0(600)}$, $m_{f_0(1370)}$, $m_{\pi_p^\prime}$ mesons and the values 
of $m_\Phi^2$, $m_{\Phi^\prime}^2$
are calculated using the extremum conditions for mesonic potential (see
eqn.~(\ref{minU})). Please note that even if $h = 0$ in this case, the
$\pi_p$ meson is still massive because of the interaction term between
quarkonium and tetraquark fields which breaks the chiral symmetry
explicitly. Since the value of the $m_{\pi}$ mass is not used in the 
parameter fixing, its mass, when calculated with the obtained parameter 
values, is comparatively higher than the real world pion mass. Since here 
we are only interested in qualitative comparison of the two scenarios and 
more elaborate studies are considered in the other cases (see Case II and 
Case III), we keep this high pion mass.

Utilizing the expressions for $\pi_p^\prime$ mass together 
with the relations:
\begin{align}
{\rm Tr}[(M_{f_0}^2)] = m_{f_0(600)}^2 + m_{f_0(1370)}^2, \\
{\rm Det}[(M_{f_0}^2)] = m_{f_0(600)}^2 \times m_{f_0(1370)}^2,
\label{scalar}
\end{align}
we get the expressions for $g_1$, $\chi$, $\lambda_1$ as:
\begin{align}
g_1 &= \frac{1}{2 f_\pi} \sqrt{\left( m_{f_0(600)}^2 + m_{f_0(1370)}^2 -
m_{\pi_p^\prime}^2 \right) m_{\pi_p^\prime}^2 - m_{f_0(600)}^2 \times
m_{f_0(1370)}^2 }, \\
\chi &= g_1 \frac{\sigma^2}{m_{\pi_p^\prime}^2}, \\
\lambda_1 &= \frac{2}{\sigma^2} \left[ m_{f_0(600)}^2 + m_{f_0(1370)}^2
- m_{\pi_p^\prime}^2 \right ].
\label{case1c}
\end{align}
Using Eq.~(\ref{minU}), we can calculate the values for $m_\Phi^2$
and $m_{\Phi^\prime}^2$ from the expressions:
\begin{align}
m_\Phi^2 &= - \left(-\frac{1}{4} \lambda_1 \sigma^2 + 2 g_1 \chi \right),
\\
m_{\Phi^\prime}^2 &= - g_1 \frac{\sigma^2}{\chi}.
\label{case1ms}
\end{align}

The value of the physical meson masses used is given in table~\ref{case1in}.
Depending on what value we choose for the mass of $m_{\pi_p^\prime}$,
we get two scenarios:

\begin{table}[htbp]
\begin{center}
\begin{tabular}{ c c c c c }
\hline \hline
Mesons & $m_{f_0(600)}$ (GeV)  & $m_{f_0(1370)}$ (GeV)
& $m_{\pi_p}$ (GeV)  & $m_{\pi_p^\prime}$ (GeV)
\\
\hline \hline
Scenario 1 & 0.8 & 1.5 & 0.42 & 1.3 \\

Scenario 2 & 0.8 & 1.5 & 0.49 & 1.1 \\
\hline \hline
\end{tabular}
\end{center}
\caption[]{Values of physical meson masses used in Case I.}
\label{case1in}
\end{table}

Scenario 1: The value of the parameter are such that the lowest scalar
$f_0(600)$ is a quarkonium dominated meson, whereas the heavier one
$f_0(1370)$ is tetraquark dominated. The value of the parameters are
given in table~\ref{case1ot}.

\begin{table}[htbp]
\begin{center}
\begin{tabular}{ c c c c c c c }
\hline \hline
Parameters & $\sigma$ (GeV) & $\chi$ (GeV)
& ${m_{\Phi}}^2$ (GeV$^2$) & ${m_{\Phi^\prime}}^2$ (GeV$^2$)
& $\lambda_1$ & $g_1$ (GeV)
\\
\hline \hline
Scenario 1 & 92.4 $\times 10^{-3}$ & 2.1 $\times 10^{-2}$ 
& 4.26 $\times 10^{-1}$ & -1.69 & 281.1 & 4.15
\\
Scenario 2 & 92.4 $\times 10^{-3}$ & 2.9 $\times 10^{-2}$  
& 5.95 $\times 10^{-1}$ & -1.21 & 393.55 & 4.17  \\
\hline \hline
\end{tabular}
\end{center}
\caption[]{Parameter set for Case I.}
\label{case1ot}
\end{table}

Scenario 2:  In this case the nature of the scalar isoscalar mesons are
 just opposite to that of Scenario 1. But for that we have to take the
input value for $m_{\pi_p^\prime}$ slightly less than its range of possible
values $1.2-1.4~{\rm GeV}$. Here, the lowest scalar
$f_0(600)$ is a tetraquark dominated meson, whereas the heavier one
$f_0(1370)$ is  quarkonium dominated. The value of the parameters so
obtained are given in table~\ref{case1ot}.

\subsection{Case II: $\lambda_2, k, h \neq 0$ but $g_2 = 0$}
To discuss how the parameter set for $g_2 = 0$ is obtained, we first
write down the equations we are going to use to determine the values
of $g_1$, $\chi$, $h$ and $k$,
\begin{align}
m_{\pi_p}^2 &= 2 \, g_1 \, \chi + 2 \, \frac{h}{\sigma},  \label{pip} \\
m_{\pi_p^\prime}^2 &= g_1 \, \frac{\sigma^2}{\chi},  \label{pip1} \\
{\rm Tr} \left[ \left(  M_\eta^2  \right) \right] &=
m_{\eta_p}^2 \, + m_{\eta_p^\prime}^2, \label{treta}
 \\
{\rm Det} \left[ \left(  M_\eta^2  \right) \right] &=
m_{\eta_p}^2 \, \times m_{\eta_p^\prime}^2.
\label{deteta}
\end{align}
Now, utilizing equations~(\ref{pip1}), (\ref{treta})
and (\ref{deteta}) we get the equation for $g_1$ as:
\begin{align}
g_1 = \frac{1}{2 \sigma} \, \sqrt{\left( m_{\eta_p}^2 \,
+ m_{\eta_p^\prime}^2 \, - m_{\pi_p^\prime}^2 \right)
 \, m_{\pi_p^\prime}^2 \, - m_{\eta_p}^2 \, \times
m_{\eta_p^\prime}^2}
\label{g1}
\end{align}
From~(\ref{pip1}), we get the vacum expectation values of the 
tetraquark field as,
\begin{align}
\chi = g_1 \, \frac{\sigma^2}{m_{\pi_p^\prime}^2}.
\label{chi}
\end{align}
Using~(\ref{pip}), (\ref{g1}) and (\ref{chi}) we can determine
the value of $h$ from the following equation,
\begin{align}
h = \frac{\sigma}{2} \, \left[ m_{\pi_p}^2 - 2 \, g_1 \,
\chi  \right].
\label{h}
\end{align}

According to convention followed in this paper the
vacuum expectation values: $\sigma$ and $\chi$ are
positive. To make sure we have the minimum of the mesonic potential
lie in the quadrant where both
$\sigma$ and $\chi$ are positive, we should have $g_1 > 0$ and
$h>0$. Apart from that we should also have positive $\lambda_1$,
$\lambda_2$ in order to make our mesonic potential bounded from below.
There is a large uncertainties in the value of
$m_{\pi_p^\prime}$~-~$(1.2-1.4)~{\rm GeV}$. But if we impose the
constraints $g_1 > 0$ and $h > 0$ and fix the values of
$m_{\eta_p}$ and $m_{\eta_p^\prime}$ at $0.55$ and $1.3~{\rm GeV}$
respectively, we find only allowed value is
$m_{\pi_p^\prime} = 1.29~{\rm GeV}$ (up to two significant
digits after decimal). Mass values higher than that would
make $g_1 < 0$ and for mass less than $1.29~{\rm GeV}$ the value
of $h$ becomes negative.

The value of $k$ can be determined from~(\ref{pip}), (\ref{pip1})
and (\ref{treta}),
\begin{equation}
k = \frac{1}{2} \, \left[ 2 \, g_1 \, \chi  \,
- \left( m_{\eta_p}^2 \, + m_{\eta_p^\prime}^2 \,
- m_{{\pi_p}^\prime}^2 - m_{\pi_p}^2 \right)  \right]
\label{k}
\end{equation}

Then the mass matrix of $f_0$ mesons can be utilized to
determine the values of $\lambda_1$ and $\lambda_2$. The
relevant equations here are:
\begin{eqnarray}
{\rm Tr} \left[ \left(  M_{f_0}^2  \right) \right] &=&
m_{f_0(600)}^2 \, + m_{f_0(1370)}^2,
\label{lambda1} \\
{\rm Det} \left[ \left(  M_{f_0}^2  \right) \right] &=&
m_{f_0(600)}^2 \, \times m_{f_0(1370)}^2.
\label{lambda2}
\end{eqnarray}

Like the value of $m_{\pi_p^\prime}$, there are also
large uncertainties in the values of
$m_{f_0(600)}$: $(0.4 - 1.2)~{\rm GeV}$ and
$m_{f_0(1370)}$: $(1.2 - 1.5)~{\rm GeV}$. Here we
have used, $m_{f_0(600)} = 0.6~{\rm GeV}$ and
$m_{f_0(1370)} = 1.35~{\rm GeV}$. Two sets of
values can be obtained from
equations~(\ref{lambda1}) and (\ref{lambda2}).
But only one of them satisfies the condition:
$\lambda_1,~\lambda_2 >0$ and considered in
this work. We have checked for other
values of $m_{f_0(600)}$ in the range
$(0.4 - 1.2~{\rm GeV})$ and $m_{f_0(1370)}$ in the
range $(1.2 - 1.5)~{\rm GeV}$ and one of the solution
for $\lambda_2$ remains always negative for the
entire mass range.

Finally the values of $m_{\Phi}^2$ and $m_{\Phi^\prime}^2$ can
be determined from the extremum condition mentioned in Eq.~(\ref{minU}).
The explicit expression for them is given below:
\begin{eqnarray}
m_{\Phi}^2 &=& - \left(-\frac{1}{4} \, \lambda_1 \, \sigma^2 + 2 \,
g_1 \, \chi - k \, + 2 \, \frac{h}{\sigma} \right),
\label{mass1} \\
m_{\Phi^\prime}^2 &=& - \left(-\frac{1}{4} \, \lambda_2 \, \chi^2 +
g_1 \, \frac{\sigma^2}{\chi} \right).
\label{mass2}
\end{eqnarray}

The values of the input physical meson masses and the resultant
output parameter set are given in Table~\ref{massT} and Table~\ref{para1}
respectively.

\begin{table}[htbp]
\begin{center}
\begin{tabular}{ c c c c c c c }
\hline \hline
Fields & $m_{f_0(600)}$ & $m_{f_0(1370)}$ & $m_{\pi_p}$ &
$m_{\pi_p^\prime}$ & $m_{\eta_p}$ & $m_{\eta_p^\prime}$
\\
\hline
Mass (GeV) & 0.6 & 1.35 & 0.14 & 1.29 & 0.55 & 1.3  \\
\hline \hline
\end{tabular}
\end{center}
\caption[]{Value of physical meson masses used for case II.}
\label{massT}
\end{table}

\begin{table}[htbp]
\begin{center}
\begin{tabular}{ c c c c c c c c c c }
\hline \hline
Parameters & $\sigma$ (GeV) & $\chi$ (GeV)
& ${m_{\Phi}}^2$ (GeV$^2$) & ${m_{\Phi^\prime}}^2$ (GeV$^2$)
& $\lambda_1$ & $\lambda_2$ & $g_1$ (GeV) & $h$ (GeV$^3$)
& $k$ (GeV$^2$)
\\
\hline \hline
Value & 92.4 $\times 10^{-3}$ & 5.23 $\times 10^{-3}$ & 1.9 $\times 10^{-2}$ 
& -1.6 & 87.99 & 9103.07
& 1.02 & 4.2 $\times 10^{-4}$ & -1.49 $\times 10^{-1}$  \\
\hline \hline
\end{tabular}
\end{center}
\caption[]{Parameter set for case II.}
\label{para1}
\end{table}

\subsection{Case III: $\lambda_2, g_2 , k, h \neq 0$ }

To fix the parameters for $g_2 \neq 0$ we follow the same procedure
as mentioned above.
Here in this case we have one more parameter $g_2$. For this,
we made an assumption that $\chi < \sigma$, which holds good for all
previous studies. Now the constraints, $g_1 > 0$, $h > 0$, $\lambda_1 > 0$,
$\lambda_2 > 0$ restrict the value of $\chi$ to a certain range. We
assume $\chi = \sigma /n$ and calculate the parameter set for small
and large possible values of "n" ($n = 10 \, \text{and} \, 18$)
respecting all the constraints.

The expressions for $g_1$, h, k remains the same as mentioned in
Eqs.~(\ref{g1}), (\ref{h}) and (\ref{k}). The expression for $g_2$
in this case reads as follows:
\begin{align}
g_2 = \frac{4}{9 \, \chi} \, \left( g_1 \, \frac{\sigma^2}{\chi} \,
- m_{\pi_p^\prime}^2 \right).
\end{align}

Here for $n = 18$, i.e., if $\chi$ is small, we get the sign of
$g_2$ to be positive. While for $n = 10$, corresponding to
comparatively large value of $\chi$, the sign of $g_2$ becomes
negative. The values of $\lambda_1$ and $\lambda_2$ are calculated using
Eqs.~(\ref{lambda1}) and (\ref{lambda2}).

Finally $m_{\Phi}^2$ and $m_{\Phi^\prime}^2$ are calculated from
the following expressions using Eq.~(\ref{minU}):
\begin{align}
m_{\Phi}^2 &= - \left(-\frac{1}{4} \, \lambda_1 \, \sigma^2 + 2 \,
g_1 \, \chi - k \, + 2 \, \frac{h}{\sigma} \right)
\label{mass11} \\
m_{\Phi^\prime}^2 &= - \left(-\frac{1}{4} \, \lambda_2 \, \chi^2 +
g_1 \, \frac{\sigma^2}{\chi} - \frac{3}{4} \, g_2 \, \chi
\right)
\label{mass22}
\end{align}

As in the case for $g_2 = 0$, here also, we only find one set of
solution which respect the constrain $\lambda_1 > 0$ and
$\lambda_2 > 0$. The values of the input physical meson masses
used here are the same as in the previous section (see Table~\ref{massT})
and the output parameters obtained are given in Table~\ref{para2}
and \ref{para3} corresponding to positive and negative $g_2$
respectively.

\begin{table}[htbp]
\begin{center}
\begin{tabular}{ c c c c c c c c c c c }
\hline \hline
Parameters & $\sigma$ (GeV) & $\chi$ (GeV)
& ${m_{\Phi}}^2$ (GeV$^2$) & ${m_{\Phi^\prime}}^2$ (GeV$^2$)
& $\lambda_1$ & $\lambda_2$ & $g_1$ (GeV) & $g_2$ (GeV) &
$h$ (GeV$^3$) & $k$ (GeV$^2$)
\\
\hline \hline
Value & 92.4 $\times 10^{-3}$ & 5.13 $\times 10^{-3}$ & 1.9 $\times 10^{-2}$ 
& -1.63 & 87.93 & 7527.9
& 10.16 $\times 10^{-1}$ & 2.25 & 4.2 $\times 10^{-4}$ & -1.49 $\times 10^{-1}$ \\
\hline \hline
\end{tabular}
\end{center}
\caption[]{Parameter set for $g_2 \neq 0$. In this set,
$\chi = \sigma/n$ where $n  = 18$ is used.}
\label{para2}
\end{table}

\begin{table}[htbp]
\begin{center}
\begin{tabular}{ c c c c c c c c c c c }
\hline \hline
Parameters & $\sigma$ (GeV) & $\chi$ (GeV)
& ${m_{\Phi}}^2$ (GeV$^2$) & ${m_{\Phi^\prime}}^2$ (GeV$^2$)
& $\lambda_1$ & $\lambda_2$ & $g_1$ (GeV) & $g_2$ (GeV) &
$h$ (GeV$^3$) & $k$ (GeV$^2$)
\\
\hline \hline
Value & 92.4 $\times 10^{-3}$ & 9.24 $\times 10^{-3}$ & 2.7 $\times 10^{-2}$ 
& -0.63 & 89.88 & 25785.1 & 1.02 & -34.88 & 3.8 $\times 10^{-5}$ 
& -1.45 $\times 10^{-1}$ \\
\hline \hline
\end{tabular}
\end{center}
\caption[]{Parameter set for $g_2 \neq 0$. In this set,
$\chi = \sigma/n$ where $n  = 10$ is used.}
\label{para3}
\end{table}


\section{Results for phase transitions}
\label{sec:phasetransition}

\begin{figure}[h]
\begin{center}
\vskip 1cm
\epsfxsize = 8.5cm
 \epsfbox{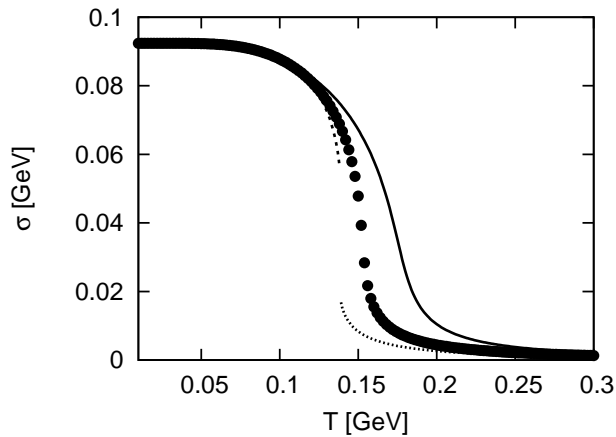}
\end{center}
\caption[]{%
Variation of $\sigma$ with temperature at zero chemical
potential for different values of $g_3$. From left to right
the values of $g_3$ are 3.5, 3.0 and 2.5. Parameters are
corresponding to the scenario $g_2 = 0$, i.e., Table~\ref{para1}.
}
\label{g3var}
\end{figure}

There are two more parameters in our model which are not discussed yet.
They are the Yukawa coupling constants $g_3$ and $g_4$. Their values
are fixed from the given value of the constituent quark mass. Since we
have one condition and two undetermined coupling constants, we assume,
$g_4 = g_3/{n_\chi}$ where ${n_\chi} > 1$. For a particular
value of $g_3$ if we change the value of $g_4$ by changing
${n_\chi}$ then there is no qualitative
change in the behaviour of $\sigma$, $\chi$. However, nature of phase
transition is affected if we change the value of $g_3$. This can
be seen from Fig.~\ref{g3var}. If we increase the value
of $g_3$ then we can get first order transition even at zero
chemical potential. This dependence of the order of the phase
transition on the values of the model parameters in mean field
approximation of Linear Sigma Model/Quark Meson Model is not new and
already noted in~\cite{heinz, Mocsy, schaefer}.
In this work, we have used $g_3 = 3.0$ and $g_4 = g_3/10$ corresponding
to vacuum constituent quark mass of 0.28 GeV. The values of $g_3$ and 
$g_4$ are so chosen to make the chiral phase transition crossover at zero 
chemical potential, as found by Lattice simulation study~\cite{fodor}.

\begin{figure}[h]
\begin{center}
\vskip 1cm
\epsfxsize = 6cm
 \epsfbox{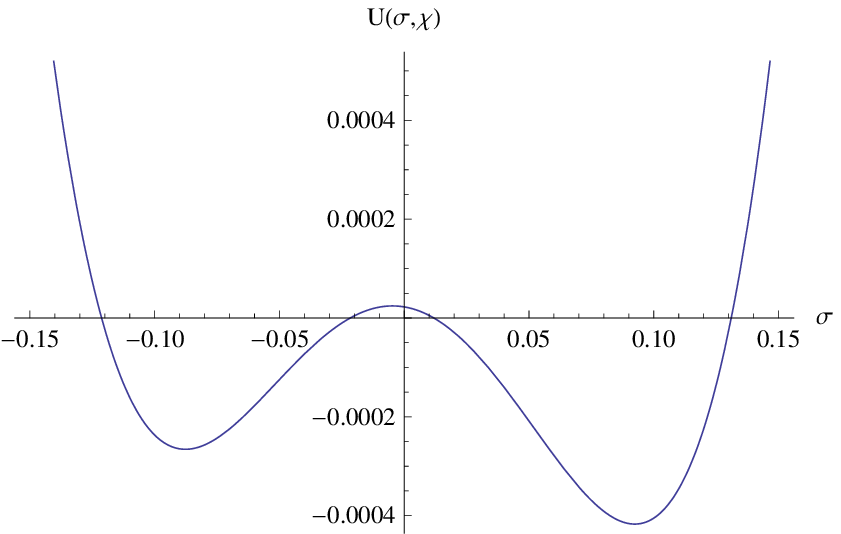}
\epsfxsize = 6cm
\epsfbox{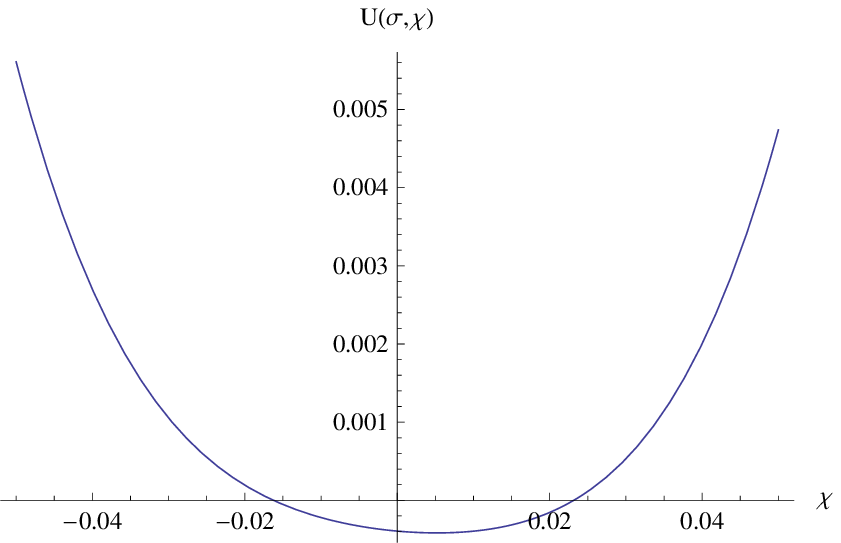}
\hskip 1cm
\end{center}
\caption[]{%
Nature of the mesonic potential $U(\sigma, \chi)$ in vacuum.
Parameters are corresponding to table~\ref{para1}. See text for
details.}
\label{potU}
\end{figure}

Before presenting our result, let us first discuss the nature of
the mesonic potential $U(\sigma, \chi)$ in vacuum. As can be seen
from the parameter set presented in~Table \ref{para1}, \ref{para2},
\ref{para3}, the sign of ${m_{\Phi^\prime}}^2$ is opposite to that
of ${m_{\Phi}}^2$. Its' sign indicates that it has opposite sign to
what requires for spontaneous breaking. This can be seen from the
figure~\ref{potU} (right one) where the potential in the $\chi$
direction (for constant $\sigma = 92.4 \times 10^{-3}$ GeV) is plotted. There is
only one minimum and the minimum of the potential is slightly tilted
in the $\chi > 0$ direction because of the explicit symmetry breaking
terms. On the other hand because of the negative sign of
${m_{\Phi}}^2$, the potential in the $\sigma$ direction exhibits the
kind of pattern expected for spontaneous symmetry breaking. The
minima in the $\sigma > 0$ direction is lower than that in the
opposite direction (see left hand figure of \ref{potU}, here
$\chi = 5.23 \times 10^{-3}$ GeV) because of $h >0$. Thus the origin of
$\sigma$ and $\chi$ condensates have different reasons in this
work. Explicit symmetry breaking is the origin for $\chi$, whereas
for $\sigma$ it is the spontaneous breaking.

For values of parameters presented in Tables~\ref{para1},
\ref{para2} and \ref{para3}, we find, irrespective of the
scalar or pseudoscalar nature of the mesons, mesons with
lower mass is always quarkonium dominated and the mesons
above 1 GeV is tetraquark dominated. The mixing angles for
$f_0$ meson for parameter sets presented in Tables~\ref{para1},
\ref{para2} and \ref{para3} are -7.51, -7.44 and -7.45 (in
degrees) respectively. For $\eta$ mesons the mixing angles
for the above mentioned parameter sets are 7.88, 7.84
and 7.86 (in degrees) respectively. Like pions, there
is no mixing for $\alpha$ meson. The masses of $\alpha_p$,
$\alpha_p^\prime$ mesons are 0.83 and 1.34 GeV respectively
for all three parameter sets. Since there is no mixing,
the lower mass $\alpha_p$ meson is purely quarkonium and
the heavier counterpart is purely tetraquark in nature.

To characterize the phase transition and to find the transition
temperature we have used the susceptibilities of the order parameters.
The susceptibility matrix is defined as \cite{harada, sasaki};
\begin{align}
\hat{\chi} = \frac{1}{C_{\chi \chi} \, C_{\sigma \sigma} \,
-C_{\sigma \chi}^2}
\left[
  \begin {array}{cc}
C_{\chi \chi}
&
C_{\sigma \chi}
\\
C_{\sigma \chi}
&
C_{\sigma \sigma}
\end{array}
 \right].
\end{align}

Where $C_{xx}$ ($x = \sigma, \chi$) are the second derivative of
the thermodynamic potential w.r.t x:
\begin{align}
C_{xx} = \frac{\partial^2 \Omega}{\partial x^2}.
\end{align}

Susceptibility of $\sigma$ is defined as $\chi_{2Q} = \hat{\chi}_{11}$
and that of $\chi$ is given by $\chi_{4Q} = \hat{\chi}_{22}$.
We determine the transition temperature from the peak position
of the respective susceptibilities. For critical point,
$C_{\chi \chi} \, C_{\sigma \sigma} \,
-C_{\sigma \chi}^2$ becomes zero corresponding to zero curvature
of the thermodynamic potential.

\subsection{Phase diagram for Case I}

\begin{figure}[h]
\begin{center}
\vskip 1cm
\epsfxsize = 7.5cm
 \epsfbox{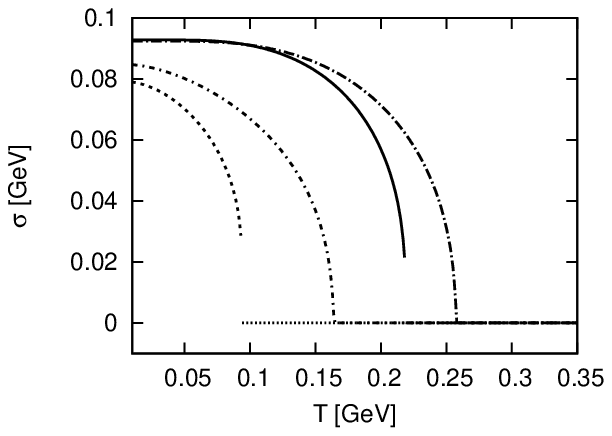}
\epsfxsize = 7.5cm
\epsfbox{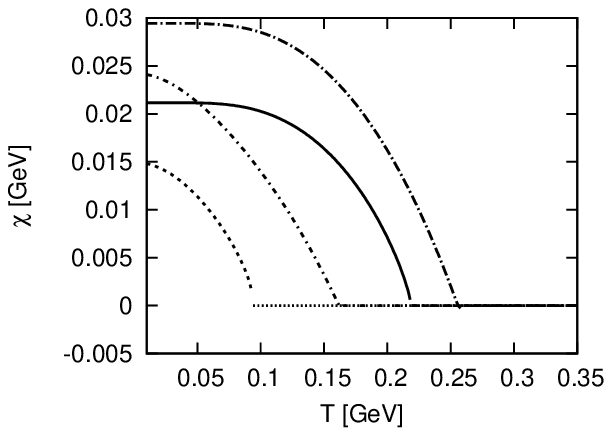}
\hskip 1cm
\end{center}
\caption[]{%
Variation of $\sigma$ and $\chi$ with temperature for different
values of chemical potential corresponding to parameter set
of Case I. The solid ($\mu = 0$ GeV) and dotted ($\mu = 0.36$ GeV )
lines are for scenario 1. Whereas, the long ($\mu = 0.0$ GeV ) and
short dash-dot ($\mu = 0.36$ GeV ) lines are for scenario 2.
}
\label{opCase1}
\end{figure}

\begin{figure}[h]
\begin{center}
\vskip 1cm
\epsfxsize = 7.5cm
 \epsfbox{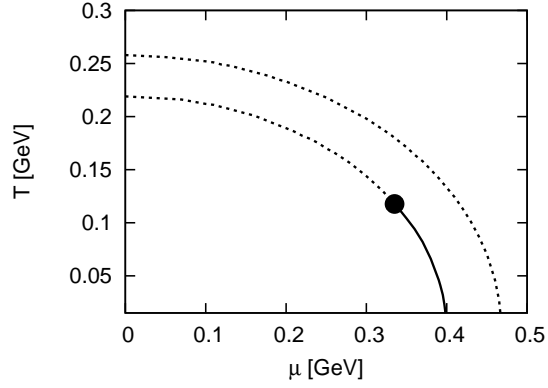}
\end{center}
\caption[]{%
Phase diagram for Case I. The dotted line represents
second order phase transition and the solid line stands
for first order phase transition. The upper phase boundary
line corresponds to scenario 2 and the lower one corresponds
to sceanrio 1.
}
\label{phdgmh0}
\end{figure}

The behaviour of the order parameters
along with the resultant
phase diagram corresponding to the parameter set for Case I
are summarized in Figs.~\ref{opCase1}, \ref{phdgmh0}.

As mentioned in the last section, here we have two scenarios
depending on the mass of $m_{\pi^\prime}$. For $m_{\pi^\prime} = 1.3$ GeV
corresponding to scenario 1, the lowest isoscalar is quarkonium
dominated. Whereas, for scenario 2, we consider $m_{\pi^\prime} = 1.1$ GeV
(which is slightly less than the value quoted in the Particle
data group: 1.2-1.4 GeV), we have lowest isoscalar as tetraquark
dominated meson.

We find, for both the cases for all values of the
chemical potential the transition temperatures calculated from the
susceptibilities $\chi_{2Q}$ and $\chi_{4Q}$ are the same. This can
also be seen from the behaviour of the order parameters presented in
Fig.~\ref{opCase1}. In Fig.~\ref{opCase1}, the temperature variation
of $\sigma$ and $\chi$ are presented for low and high values of chemical
potential. From the figure, we see $\sigma$ and $\chi$ varies more
slowly with temperature in the case for scenario 2 than in scenario 1.
Consequently, the transition temperature in scenario 2 is always higher
than the scenario 1. For both the scenarios, both $\sigma$ and $\chi$
goes to zero after the phase transition because of $h = 0$. But for
scenario 1, there is a jump in case of $\sigma$ after a certain
temperature and this gap in the order parameter increases slowly with
the chemical potential. For $\chi$, this gap is vanishingly small
at low chemical potential and slowly increases with the chemical potential.

If we compare the phase diagrams shown in Fig. (\ref{phdgmh0}), we see for
scenario 2, the order of the phase transition is second order both
for low as well as high values of the chemical potential. But for
scenario 1, the second order phase transition changes to weak
first order phase transition above some critical value of the chemical
potential, thus indicates presence of a critical point. We consider the
values of the chemical potential and temperature at which the curvature of
the thermodynamic potential becomes greater than $10^{-4}$,
as the location of the critical point. Using this condition, we find
the critical point for scenario 1 is located at $T_c = 117.7$ MeV and
$\mu_c = 335$ MeV. The departure from the zero curvature of the
thermodynamical potential together with the gap in the order parameter
are taken as the indication of weak first order phase transition.
We are calling it weak first order because curvature of the thermodynamic
potential remains very small ($ \sim 10^{-3}$) for $\mu > 335$  MeV.

\subsection{Phase diagram for Case II and Case III}

\begin{figure}[h]
\begin{center}
\vskip 1cm
\epsfxsize = 7.5cm
 \epsfbox{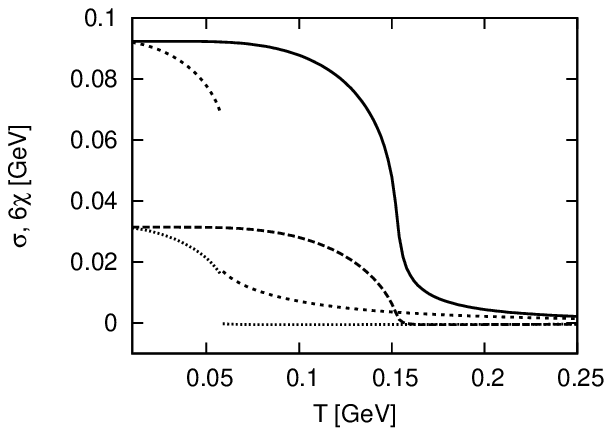}
\epsfxsize = 7.5cm
\epsfbox{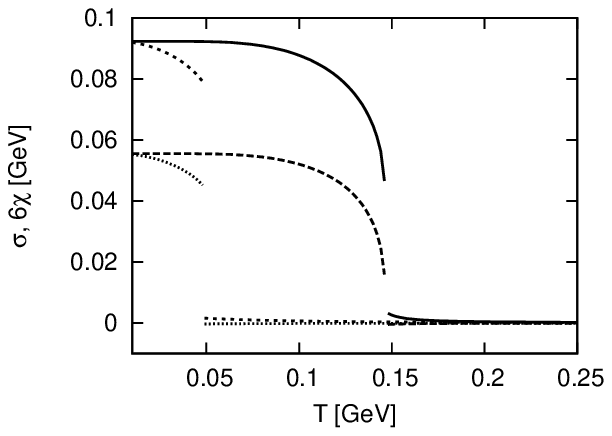}
\hskip 1cm
\end{center}
\caption[]{%
Variation of $\sigma$ and $\chi$ with temperature for different
values of chemical potential. Figure in the left panel is for
$g_2 = 0$ and the right one is for $g_2 = -34.88$. The
solid ($\mu = 0$ GeV) and short dash ($\mu = 0.27$ GeV) lines
are for variation of $\sigma$. Variation of $\chi$ is represented
by long dash ($\mu = 0$ GeV) and points ($\mu = 0.27$ GeV) for
both the figures.
}
\label{op}
\end{figure}

\begin{figure}[h]
\begin{center}
\vskip 1cm
\epsfxsize = 7.5cm
\epsfbox{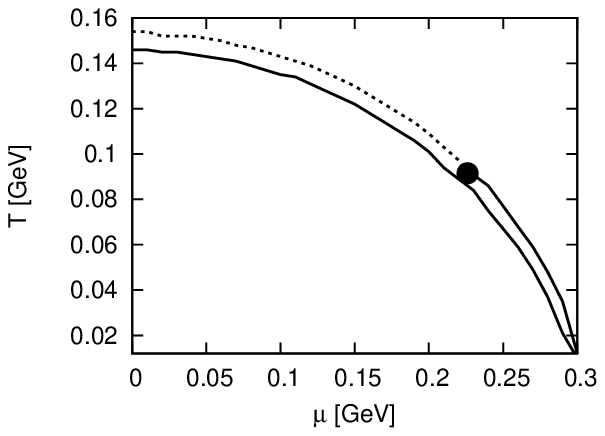}
\hskip 1cm
\end{center}
\caption[]{%
Phase diagram for Case III. The solid line indicates first
order phase transition and the dashed line is for crossover
transition. The upper phase boundary is for $g_2 = 2.25$ and
the lower one is for $g_2 = -34.88$. The bold circle indicates
location of the critical end point. See text for details}
\label{phdgm}
\end{figure}

The nature of the phase transition corresponding to Case II and III
are summarized in Figs.~\ref{op} and \ref{phdgm}.

Behaviour of the order parameters at low and high values of the
chemical potentials are presented in Fig.~\ref{op}, where the
left figure corresponds to $g_2 = 0$ and the right one $g_2 < 0$.
We find for $g_2 = 0$ (see Table~\ref{para1}) and
$g_2 > 0$ (see table~\ref{para3}) the behaviour of the order
parameters are qualitatively similar. This is expected as the
positive cubic interaction coupling constant for the tetraquark field 
is relatively small and the other parameters being almost of same values. As can be
seen from Fig. ~\ref{op} (left), for small chemical potential
we have a crossover transition which turns into first order
transition at high value of the chemical potential. Because of finite
"$h$" term for the $\Phi$ field makes $\sigma > 0$ even at high temperature.
But absence of such term for the $\Phi^\prime$ field makes $\chi$ goes
to zero at high temperature. However, the nature of the transition is
quite different corresponding to scenario $g_2 < 0$ (see Table~\ref{para2}).
In this case, the strong cubic interaction term make the transition
first order for the whole range of chemical potential as can be seen
from Fig.~\ref{op} (right). Here relatively low value of "$h$"
makes $\sigma$ goes to zero at high temperature. However, there is
one similarity with respect to chiral phase transition temperature
for various cases considered in this work. Like Case I, we note
from both the figures in Fig.~\ref{op}, the transition for $\sigma$
and $\chi$ are occurring at the same temperature which is verified
from the peak positions of the respective susceptibilities.

As a result, the resultant phase diagram shown in Fig.~\ref{phdgm}
is represented with a single phase boundary line for each case. For
$g_2 = 0$ and $g_2 > 0$ we have qualitatively same feature and thus we have
included the phase boundary for Case III only in Fig.~\ref{phdgm}.
In this case, we have a crossover transition at low chemical potential
which turns into first order above some critical value of the chemical
potential. Thus, we have a critical end point for $g_2 = 0$ and $g_2 > 0$.
The location of the critical end point for $g_2 = 0$ is ($\mu = 0.26$ GeV,
$T = 0.069$ GeV) and for $g_2 > 0$ is ($\mu = 0.226$ GeV, $T = 0.0915$ GeV).
Whereas, for $g_2 < 0$ we have only first order phase transition line owing
to strong cubic interaction term.

\section{Summary and Conclusion}

In the framework of two flavor quark-meson model, we have
investigated the effect of mixing between quarkonium and
tetraquark fields on chiral phase transition.

The mixing between the effective fields is introduced through an
interaction term which breaks the chiral symmetry explicitly.
In addition to the interaction term, we also considered a
cubic self interaction term for the effective tetraquark
field, an instanton determinant term and a term mimicking
the effect of current quark mass. The parameters of our model
is calculated from the masses of the physical mesons, pion
decay constants and the stability conditions of the mesonic
potential. We first considered the effect of the mixing term
without considering the cubic self interaction term for the
tetraquark field, the term mimicking the current quark mass and
the instanton determinant term. Within the allowed experimental
range for the masses for $f_0(600)$, $f_0(1370)$, $\pi$ and
$\pi^\prime$ mesons we find our lowest scalar $f_0(600)$ meson
is quarkonium dominated. 


For the scenario where $f_0(600)$ is tetraquark dominated, we
find the chiral phase transition is second order for both low
and high values of the quark chemical potential. On the other
hand, if we increase the absolute value of the mass of the bare
tetraquark field, thereby increasing the value of the $\pi^\prime$
mass, we can have a weak first order phase transition above some
critical value of the chemical potential. Comparing the transition
in both cases, we find transition temperature is lowered with the
increase of the absolute value of bare tetraquark field mass.

Next we study the effect of the cubic self interaction term
(with coupling constant $g_2$) of the tetraquark fields. We find
physical meson mass
spectrum and the vacuum stability conditions put a tight constraint
on our parameter set. From the resulting parameter sets, we find
lowest scalar meson $f_0(600)$ is a quarkonium dominated meson
whereas, $f_0(1370)$ is tetraquark dominated. For $g_2 = 0$
(but including the effect of finite current quark mass and the
instanton term) and small but positive $g_2$, the chiral phase
transition is crossover for small values of the quark chemical
potential and then above some critical value of the chemical
potential the transition becomes first order. Thus we have a
critical end point in this case and the resultant phase diagram
matches well with current consensus regarding the two flavor phase
diagram. But with a strong and negative $g_2$ makes not only the
transition of $\chi$ first order but the transition for $\sigma$ as
well becomes first order irrespective of the low or high value of
the quark chemical potential. The strong and negative $g_2$ also makes
the chiral phase transition temperature lowered than that for the case
of $g_2 = 0$ or positive. For all the various scenarios considered in
our study, the common feature among all of them is that the transition
for quarkonium and tetraquark happening at the same temperature for all
values of the chemical potential.

\hskip 1cm

\noindent
{\bf Acknowledgments:}\\
This work is supported by the NSFC under Grants
No. 11250110058 and No. 11275213,  DFG and NSFC (CRC 110), CAS fellowship
for young foreign scientists under
Grant No. 2011Y2JB05, CAS key project KJCX2-EW-N01, K.C.Wong
Education Foundation and Youth Innovation Promotion Association of CAS.


\end{document}